\documentclass[]{spie}  

\usepackage[]{graphicx}
\usepackage[]{epstopdf}
\usepackage[]{bigints}
\usepackage{epsf}
\title{Speckle correction in polychromatic light with the self-coherent camera for the direct detection of exoplanets}

\author{J.~Mazoyer\supit{a}, R.~Galicher\supit{a}, P.~Baudoz\supit{a}, G.~Rousset\supit{a}
\skiplinehalf
\supit{a} LESIA, Observatoire de Paris, CNRS, UPMC Paris 6 and Denis Diderot Paris 7, Meudon, France
}

\authorinfo{Further author information: (Send correspondence to J.M.)\\J.M.: E-mail: johan.mazoyer@obspm.fr, Telephone: 33 6 45 07 75 12}

\begin{document} 
  \maketitle 

\begin{abstract}
Direct detection is a very promising field in exoplanet science. It allows the detection of companions with large separation and allows their spectral analysis. A few planets have already been detected and are under spectral analysis. But the full spectral characterization of smaller and colder planets requires higher contrast levels over large spectral bandwidths. Coronagraphs can be used to reach these contrasts, but their efficiency is limited by wavefront aberrations. These deformations induce speckles, star lights leaks, in the focal plane after the coronagraph. The wavefront aberrations should be estimated directly in the science image to avoid usual limitations by differential aberrations in classical adaptive optics. In this context, we introduce the Self-Coherent Camera (SCC). The SCC uses the coherence of the star light to produce a spatial modulation of the speckles in the focal plane and estimate the associated electric complex field. Controlling the wavefront with a deformable mirror, high contrasts have already been reached in monochromatic light with this technique. The performance of the current version of the SCC is limited when widening the spectral bandwidth. We will present a theoretical analysis of these issues and their possible solution. Finally, we will present test bench performance in polychromatic light.\end{abstract}


\section{Introduction}
\label{sec:intro}  

The detection and analysis of exoplanets is one of the major challenge in today's astronomy. Most of the known exoplanets were found using indirect methods. However, direct imaging of exoplanets is crucial to extend our knowledge of these objects. First, it allows the detection of long orbit objects. In addition, the analysis of these planet using spectroscopic analysis is now possible, and is necessary for the study of atmospheres and surfaces (chemistry, temperature, habitability). A few planets\cite{Marois06, Lagrange_Bpic09} have already been detected using this method and are under analysis. 

However, a lot of detections are prevented by the high contrast between the planet and its star. The required contrast levels go from $10^{-6}$ for young Jupiter-like planets to $10^{-10}$ for rocky planets. To reach these contrast levels in the focal plane of a telescope, one can use coronagraphy. This method is using  a mask in focal plane to diffract the on-axis light (the star). This diffracted light is then stopped in the next pupil plane by a Lyot stop. On the contrary, off-axis light (planet) remain unchanged. Several missions under development plan to use coronagraphy to detect exoplanets with ground based telescopes (SPHERE\cite{Beuzit08} and GPI\cite{Macintosh08}) or space-based telescopes (NIRCAM and MIRI\cite{Boccaletti05_MIRI}, aboard the JWST). But coronagraph techniques are limited by the wavefront aberrations introduced by the atmosphere or by the optics encountered. These aberrations produce speckles, remains of the stars light, in the detector focal plane, which limit the contrast and thus prevent the planet detection. 

To correct these aberrations, one must be able to measure them directly in the science focal plane, avoiding the passage through an analysis channel, which introduce differential aberrations. Several techniques have been developed to perform this focal plane analysis. We present here one of this technique, the self-coherent camera\cite{Baudoz06,Galicher10}. This instrument uses the coherence of the star light to encode the speckles and retrieve the complex speckle field in the focal plane. Associated with a deformable mirror (DM), it allows the correction of speckles in closed-loop. 

After recalling the principle of the self-coherent camera (SCC), we will present the latest results obtained on the experimental test bench in monochromatic and in polychromatic light. 

\section{Self-coherent-Camera : principle} 
\label{sec:SCCprinc}

\subsection{Speckle complex field estimation in the focal plane downstream of the coronagraph} 	
\label{sec:specle_estimation}

 \begin{figure}
 \begin{center}
 \begin{tabular}{c}
 \includegraphics[height=5.5cm]{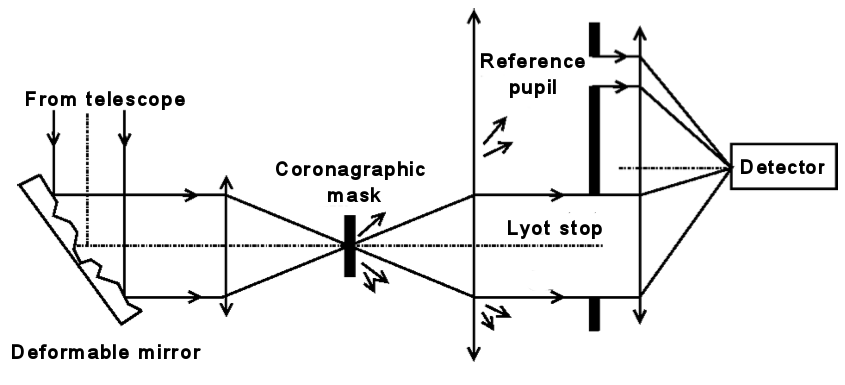}
 \includegraphics[height=5cm]{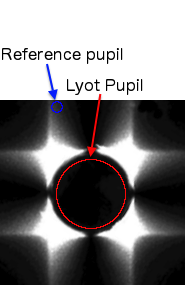}
 \end{tabular}
 \end{center}
 \caption[scc_princ] 
{ \label{fig:scc_princ} 
Principle of the SCC associated with a coronagraph and a DM (left). On the right, light distribution of a four quadrant phase mask (FQPM) coronagraph in the Lyot plane, with the position of the Lyot pupil (red) and reference pupil (blue).}
   \end{figure} 

\begin{figure}
 \begin{center}
 \begin{tabular}{c}
 \includegraphics[height=5cm]{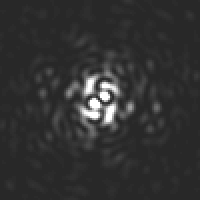}
 \includegraphics[height=5cm]{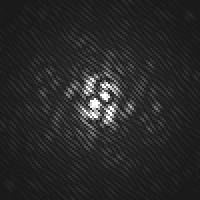}
 \includegraphics[height=5cm]{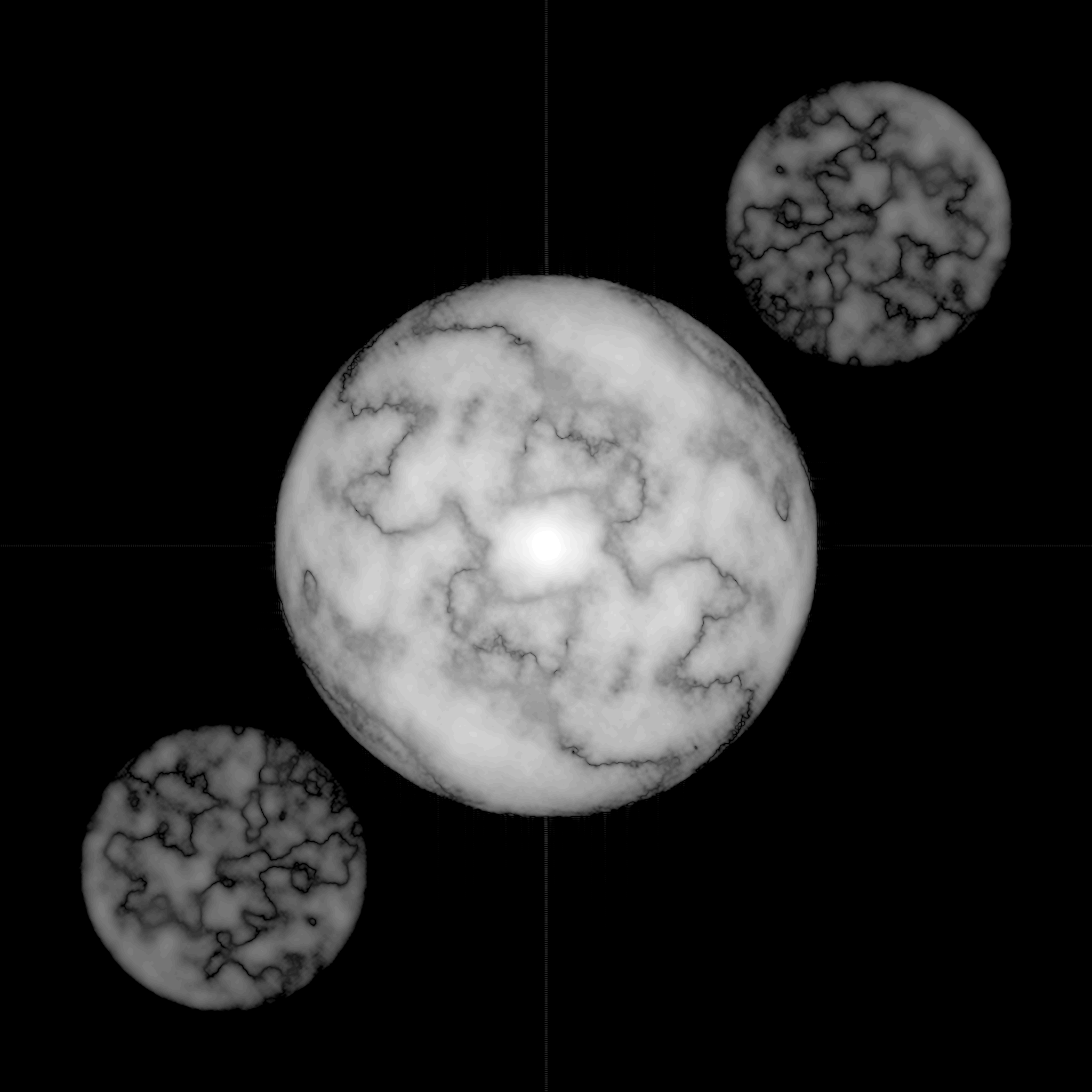}
 \end{tabular}
 \end{center}
 \caption[fringes] 
{ \label{fig:fringes} 
Simulated speckles field in the focal plane of a coronagraph (left). The same speckles field, spatially modulated (center) by the SCC. On the right, the Fourier transform of this focal plane.}
   \end{figure}

Coronagraphs have proven their efficiency to reach high contrast levels. After an entrance pupil (of size $D_P$), the stellar light is focalized on a mask, which diffracts the on-axis light. In a following pupil plane, a Lyot pupil (of size $D_L$ slightly smaller than $D_P$) removes this diffracted light. We then focalize the light on the detector. However, aberrations in the incoming wavefront result in stellar leaks or speckles in the Lyot stop (Figure~\ref{fig:fringes}, left). The SCC requires a small modification of the coronagraph, as shown in Figure \ref{fig:scc_princ} (left). In the Lyot stop plane of a coronagraph, we add a small pupil (of size $D_R$, smaller than $D_L$), to which we will refer as \emph{reference pupil} (Figure \ref{fig:scc_princ}, right). This allows us to collect some of the light diffracted by the coronagraphic mask and produces fringes in the focal plane. In Figure~\ref{fig:fringes} (left), we represent the simulated speckles field in a classical coronagraph. In Figure~\ref{fig:fringes} (center), the same image after the addition of the reference pupil. We will refer to this image as \emph{SCC image}. 

In this section, we will describe the SCC used in monochromatic light. As described in Galicher et al. 2010\cite{Galicher10}, the image in the detector plane, in the absence of noise and in monochromatic light, can be written: 
\begin{equation}
	\label{eq:I_monoc_1ref}
	\begin{array}{c}
\ I(\vec{\alpha},\lambda_0) = |A_{S}(\vec{\alpha},\lambda_0)|^2 + |A_{R}(\vec{\alpha},\lambda_0)|^2 \\
\ + A_{S}(\vec{\alpha},\lambda_0)A_{R}^{*}(\vec{\alpha},\lambda_0)\exp\left( \frac{2i\pi\vec{\alpha}.\vec{\xi_0}}{\lambda_{0}} \right) + A_{S}^{*}(\vec{\alpha},\lambda_0)A_{R}(\vec{\alpha},\lambda_0) \exp\left( -\frac{2i\pi\vec{\alpha}.\vec{\xi_0}}{\lambda_{0}} \right),
 \end{array}
	\end{equation}
where $A_{S}(\vec{\alpha},\lambda_0)$ (respectively $A_{R}(\vec{\alpha},\lambda_0)$) is the complex amplitude in the focal plane of the speckles (respectively of the reference pupil), $\vec{\alpha}$ is the focal plane angular coordinate, and $\vec{\xi_{0}}$ is the position of the center of the reference pupil with respect to the center of the Lyot pupil in the Lyot stop plane.

The Fourier transform ($\mathcal{F}$) is therefore composed of three distinct peaks. We decompose our SCC image into the sum of three images: $I = I_{cent} + I_{+} + I_{-}$ which verify 
\begin{equation}
	\label{eq:Icent}
\mathcal{F}[I_{cent}] = \mathcal{F}[|A_{S}|^2 + |A_{R}|^2]
	\end{equation}
for the central peak and
\begin{equation}
	\label{eq:I+and-}
	\begin{array}{c}
\mathcal{F}[I_{-}](\vec{u}) = \mathcal{F}[A_{S}A_{R}^{*}](u)\ast\delta(\vec{u} - \vec{\xi_0}/\lambda_0)\\
\mathcal{F}[I_{+}](\vec{u}) = \mathcal{F}[A_{S}^{*}A_{R}](u)\ast\delta(\vec{u} + \vec{\vec{\xi_0}}/\lambda_0)
 \end{array}
	\end{equation}
for the lateral ones (where $\vec{u}$ is the coordinate in the Fourier plane and $\delta$ the Kronecker delta). The central peak is of diameter $2D_{L}/\lambda_0$ and the two lateral peaks are of diameter $d_{peaks} = (D_{L}+D_{R})/\lambda_0$, and at a distance $l_{peaks} = \Vert\vec{\xi_0}\Vert/\lambda_0$ from the center. The Fourier transform of a simulated SCC image is represented in Figure \ref{fig:fringes} (right).

With an inverse Fourier transform of the lateral peak, we gain access to $A_{S}A_{R}^{*}$. From this estimate, we developed two techniques to correct speckles. In the first one, developed in Mazoyer et al. (2013)\cite{mazoyer13}, we use a model of the coronagraph to produce an estimate of the phase and amplitude defects in the entrance pupil before the coronagraph. Using the DM, we are able to correct for the phase using adaptive optic techniques. The minimization of the phase resulted in the correction of the speckles in the focal plane. In the second method, we directly minimize $A_{S}A_{R}^{*}$, the amplitude of the speckles in focal plane using an interaction matrix. This method will be detailed in the next section.

\subsection{Interaction matrix and correction} 	
\label{sec:matrix}

\begin{figure}
 \begin{center}
 \begin{tabular}{c}
 \includegraphics[height=4cm]{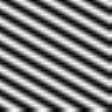}
 \includegraphics[height=5.5cm]{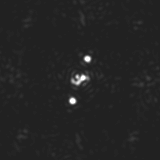}
 \includegraphics[height=5.5cm]{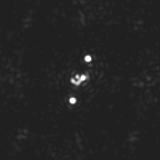}
 \end{tabular}
 \end{center}
 \caption[Iplusmoinsetdiff] 
{ \label{fig:Iplusmoinsetdiff} 
On the left, the voltage map applied apply to the 32x32 actuators of the DM to realize one of the cosinus functions on the DM. Every pixel is a voltage to apply to a actuators. We reccord the images in focal plane of the SCC when we apply to the DM's actuators the vector of tension $V_{ini} + V_{cos}$ (center) and when we apply the vector of tension $V_{ini} - V_{cos}$ (right).}\end{figure}

\begin{figure}
 \begin{center}
 \begin{tabular}{c}
 \includegraphics[height=5.5cm]{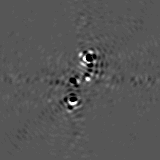}
 \includegraphics[height=5.5cm]{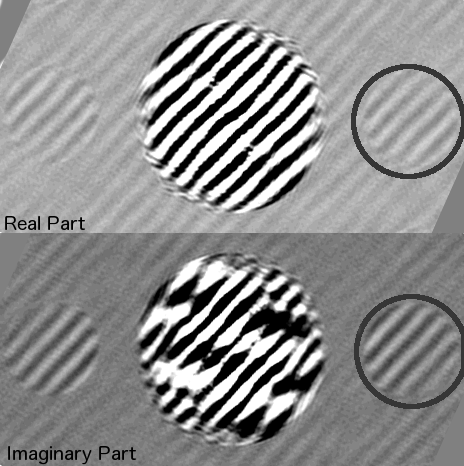}
 \includegraphics[height=5.5cm]{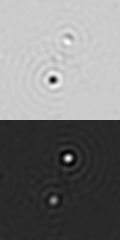}
 \end{tabular}
 \end{center}
 \caption[fft_diff_et_ASEstim] 
{ \label{fig:fft_diff_et_ASEstim} 
On the left image, we subtract the two SCC images obtain in Figure \ref{fig:Iplusmoinsetdiff} (center and right) to limit the influence of other aberrations. We apply a Fourier transform to the difference of images (center). The real and imaginary part of this Fourier transform are represented here on the left. We isolate the lateral peak ($\mathcal{F}[A_{S}A_{R}^{*}](u)\ast\delta(\vec{u} - \vec{\xi_0}/\lambda_0)$), center it ($\mathcal{F}[A_{S}A_{R}^{*}](\vec{u})$) and apply an inverse Fourier transform ($A_{S}A_{R}^{*}$). The imaginary and real parts of the result of this operation is represented on the right.}
\end{figure}

In this section, we explain how we create an interaction matrix linking the voltages applied to the actuators of the DM to their influence in the focal plane using our estimate of the complex electric field of the focal plane. Figure~\ref{fig:Iplusmoinsetdiff}~and~\ref{fig:fft_diff_et_ASEstim} present experimental images conducted on a optical bench of the creation of a matrix.

We apply on the actuators mirror 1024 cosinus and sinus functions of known frequencies. In Figure \ref{fig:Iplusmoinsetdiff} (left), we represent one of the voltage vector $V_{cos}$ applied on the 32x32 actuators of our mirror. Starting with a given voltage applied $V_{ini}$ to our DM's actuators, we take two SCC images, one when applying $V_{ini} + V_{cos}$ to the mirror ($I^{ini+cos}$, represented on Figure \ref{fig:Iplusmoinsetdiff}, center) and one when applying $V_{ini} - V_{cos}$ ($I^{ini - cos}$, represented on Figure \ref{fig:Iplusmoinsetdiff}, right). We take the difference of these two images (\ref{fig:fft_diff_et_ASEstim}, left) to get rid of the initial aberrations that we want to correct: assuming small aberrations, we use a linear model and write:

\begin{equation}
	\label{eq:difference}
I^{ini+cos} - I^{ini-cos} = 2I^{cos}.
	\end{equation}

The reason of the choice of this basis of actuator movements (sinus and cosinus) appears in Figure \ref{fig:Iplusmoinsetdiff}. Each cosinus (or sinus) applied on the DM will have a very localized influence in the focal plane: two speckles, whose positions are determine by the frequency of the applied cosinus (or sinus). We apply a Fourier transform. The real and imaginary part of this Fourier transform are represented in Figure \ref{fig:fft_diff_et_ASEstim} (center). We select the lateral peak (equal to $\mathcal{F}[A_{S}A_{R}^{*}](u)\ast\delta(u - \vec{\xi_0}/\lambda_0)$) and center it (then equal to $\mathcal{F}[A_{S}A_{R}^{*}](u)$). Finally, we apply an inverse Fourier transform (we get $A_{S}A_{R}^{*}$). The result is a 2D complex function of size NxN, whose real and imaginary part are represented in Figure~\ref{fig:fft_diff_et_ASEstim} (right). The real and imaginary parts of this 2D function are then reform into one 1D vector of size $2N^2$. We iterate for every frequencies accessible to our DM. We create this way an interaction matrix of size 1024x$2N^2$ that we can invert to produce a control matrix. This control matrix will allow the correction of an unknown wavefront in closed loop. We applied the closed-loop for a few iterations (typically a dozen) and get a stable correction. Because a NxN DM can only reach a limited number of frequencies, the corrected zone will be limited in the focal plane to a square of $N\lambda_0/D$x$N\lambda_0/D$, called the dark hole (DH). The results of this correction will be presented in Section \ref{sec:results}.

In the next section, we develop the issues raising in polychromatic light. We will then detailed the improvements made to this method to take them into account.

\subsection{Polychromatic light} 	
\label{sec:polychromatic}

\subsubsection{Differences in the estimation} 	
\label{sec:polychromatic_light_theo}
For a larger bandwidth $[\lambda_{min}, \lambda_{max}]$, we define $r_{\lambda} = \lambda /(\lambda_{max} -\lambda_{min})$ the spectral bandwidth and $\lambda_0 = (\lambda_{max} +\lambda_{min})/2$ the central wavelength. We assume that all the aberrations are achromatic, meaning that they produce the same speckles in a focal plane at every wavelengths, with only a dilatation: 

\begin{equation}
	\label{eq:dilate}
	\begin{array}{c}
A_{S}(\vec{\alpha},\lambda) = A_{S}\left(\frac{\vec{\alpha}\lambda}{D}\right)\\
A_{R}(\vec{\alpha},\lambda) = A_{R}\left(\frac{\vec{\alpha}\lambda}{D}\right)
  \end{array}
	\end{equation}

The three terms introduced in Equation \ref{eq:I_monoc_1ref}, \ref{eq:Icent} and \ref{eq:I+and-} becomes:
\begin{equation}
	\label{eq:I_polyc_1ref}
	\begin{array}{c}
I_{cent}(\vec{\alpha}) =  \bigintss\limits_{\lambda_{min}}^{\lambda_{max}}\big|A_{S}\left(\frac{\vec{\alpha}\lambda}{D}\right)\big|^2 + \big|A_{R}\left(\frac{\vec{\alpha}\lambda}{D}\right)\big|^2 \mathrm{d}\lambda,\\
I_- =  \bigintss\limits_{\lambda_{min}}^{\lambda_{max}}A_{S}\left(\frac{\vec{\alpha}\lambda}{D}\right)A_{R}^{*}\left(\frac{\vec{\alpha}\lambda}{D}\right)\exp\left( \frac{2i\pi\vec{\alpha}\xi_{0}}{\lambda} \right)\mathrm{d}\lambda,\\
I_+ =   \bigintss\limits_{\lambda_{min}}^{\lambda_{max}}A_{S}^{*}\left(\frac{\vec{\alpha}\lambda}{D}\right)A_{R}\left(\frac{\vec{\alpha}\lambda}{D}\right) \exp\left( -\frac{2i\pi\vec{\alpha}\xi_{0}}{\lambda} \right)\mathrm{d}\lambda.
  \end{array}
	\end{equation}

\begin{figure}
 \begin{center}
 \begin{tabular}{c}
 \includegraphics[height=6cm]{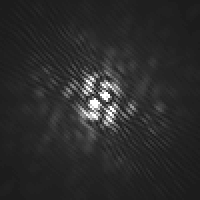}
 \includegraphics[height=6cm]{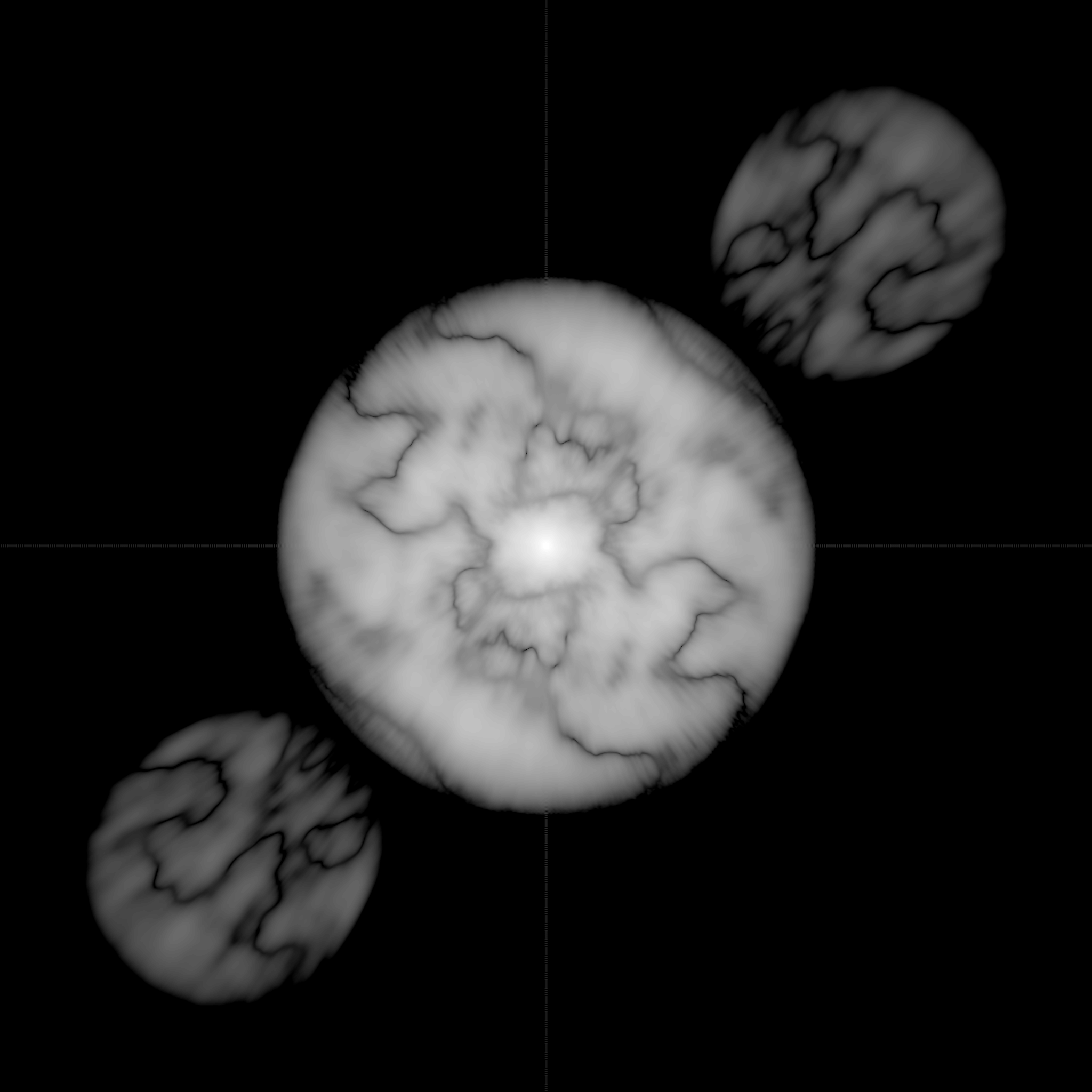}
 \includegraphics[height=6cm]{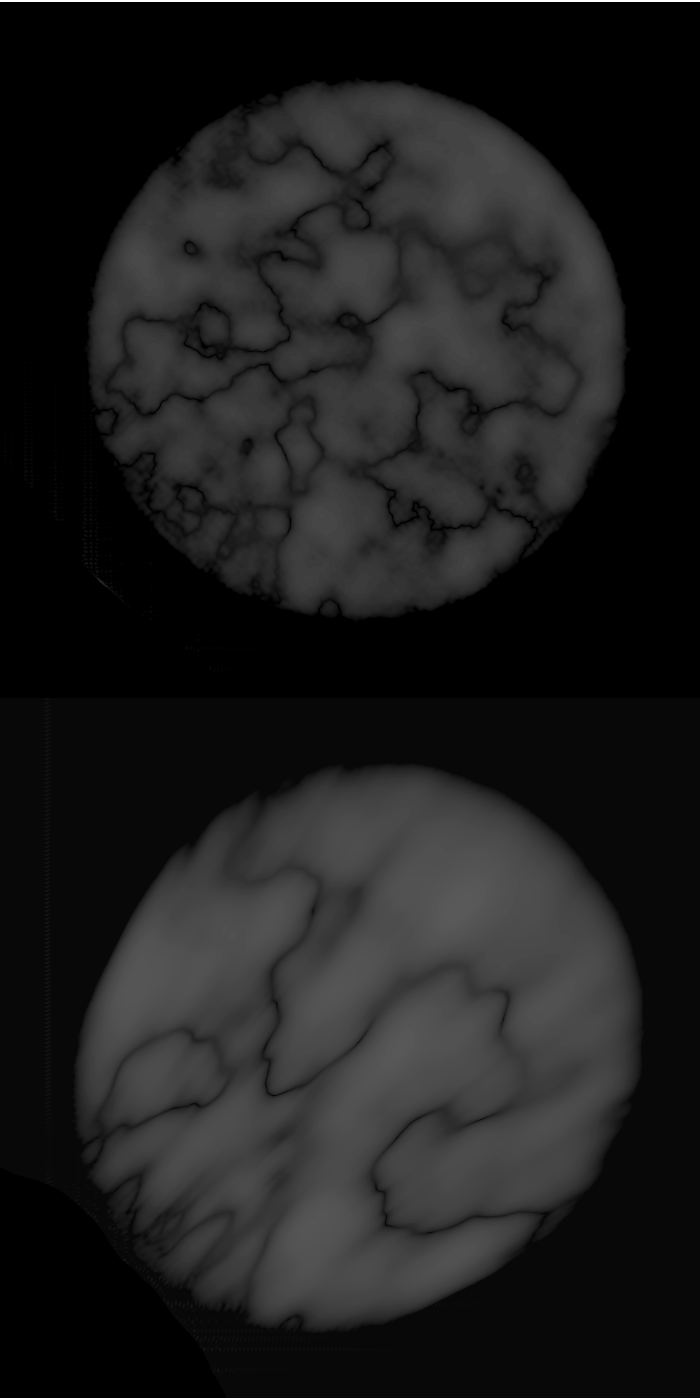}
 \end{tabular}
 \end{center}
 \caption[difference_in_polychrom] 
{ \label{fig:difference_in_polychrom} 
Simulated speckles field (right) in the focal plane of the SCC in polychromatic light ($r_\lambda = 8$). Due to the multiple wavelengths, the fringes are contrasted in one direction (the orthogonal direction of the line joining the center of the Lyot pupil to the reference pupil) and blurred in the other one. We apply a Fourier transform to this image (center). An other effect of the multiple wavelength is the deformation of the lateral peaks $\mathcal{I_{-}}$ and $\mathcal{I_{+}}$. This deformations can be seen of the right image, where we show a zoom on the lateral peaks, in monochromatic light (right, up) and in polychromatic light (right, bottom) for $r_\lambda = 8$.}
   \end{figure}

When using the SCC with monochromatic light, the superposition of multiple fringe patterns with different wavelength (and thus different inter-fringes) will tend to blur the fringes far away from the white fringe (null optical path distance). This effect was already described in Galicher et al. 2010\cite{Galicher10}. Therefore, the focal plane will only be fringed in a stripe of width $d_{r_{\lambda}}$. We can consider that the fringes are completely blurred when the two fringe patterns at $\lambda_{min}$ and $\lambda_{max}$ are shifted from half an inter-fringes. We deduce\cite{Galicher10}:
\begin{equation}
	\label{eq:interfringes}
d_{r_{\lambda}} = \frac{r_{\lambda}\lambda_0}{2\vec{\xi_0}}
	\end{equation}
Figure~\ref{fig:difference_in_polychrom} (left) shows this stripe in a speckles fringes for $r_\lambda = 8$ (in this simulation the null optical path distance goes by the center of the image). This effect is important as we can only estimate $A_S$ in the fringed zone. For a NxN DM, the maximum DH size is $N\lambda_{min}/D$, but a large bandwidth can limit the correction to a smaller zone. In the next section, we introduce an other method to correct for in a larger zone.

One can notice that the same effect of superposition of wavelengths also appears in the Fourier plane. The lateral peaks  $\mathcal{F}[I_{+}]$ now reads:
\begin{equation}
	\label{eq:lat_peak_poly}
\mathcal{F}[I_{+}](\vec{u}) = \bigintss\limits_{\lambda_{min}}^{\lambda_{max}}\left(\frac{D}{\lambda}\right)^2\mathcal{F}[A_{S}^{*}A_{R}](\vec{u}\lambda/D)\ast\delta(\vec{u} - \vec{\xi_0}/\lambda)\mathrm{d}\lambda,
	\end{equation}
thus is a superposition of peaks situated of size $d_{peaks} \in [(D_{L}+D_{R})/\lambda_{max}, (D_{L}+D_{R})/\lambda_{min}]$ and of distance to the center $l_{peaks} \in [\Vert\vec{\xi_0}\Vert/\lambda_{max}, \Vert\vec{\xi_0}\Vert/\lambda_{min}]$. The relative intensity of these peaks decreases as they are closer to the center. The Fourier transform of a polychromatic SCC image for $r_\lambda = 8$ in shown in Figure \ref{fig:difference_in_polychrom} (center). We zoom on the lateral peaks created in monochromatic light (right, up) and in polychromatic light (right, bottom), for the same wavefront aberrations. Therefore, at large bandwidth, the lateral peak has to be carefully selected. 

\subsubsection{Introduction of a second reference pupil} 	
\label{sec:2refs}

To improve the performance in polychromatic light, an upgrade of the SCC is developed. Indeed, we saw that the fringes get blurred in one direction and thus, the speckles cannot be suppressed there. The addition of a second reference pupil of the same size but at a different angle in the Lyot plane produces an other fringe pattern contrasted in a different direction. It expands the fringed zone and provides a more complete estimate of the complex field in the focal plane. Therefore, it enlarges the correction zone for large bandwidth. A study of the multi-reference SCC including experimental results will be presented in a forthcoming paper. 

In the next section, we describe the components of the test bench. Results in monochromatic and polychromatic lights are presented in Section~\ref{sec:monochromatic_res}~and~\ref{sec:polychromatic_res}.

\section{Bench description} 
\label{sec:bench_desc}

The THD test bench is located in Paris Observatory, Meudon, France. We present the main components below:

\begin{enumerate}
\item An optical fiber source fed either by a quasi-monochromatic fiber laser diode at 637 nm ($\Delta \lambda < 1$ nm) or by a super-continuum fiber laser source.
\item A set of four filters to select the central wavelength $\lambda_0$ and spectral bandwidth $\delta \lambda$. We can choose between:  
\begin{itemize}
\item[(a)]$\lambda_0 = 615.1$ nm and $\Delta \lambda = 8.6$ nm,
\item[(b)]$\lambda_0 = 633.2$ nm and $\Delta \lambda = 7.5$ nm,
\item[(c)]$\lambda_0 = 637.4$ nm and $\Delta \lambda = 9.9$ nm,
\item[(d)]$\lambda_0 = 643.7$ nm and $\Delta \lambda = 9.4$ nm,
\item[(e)]$\lambda_0 = 652.2$ nm and $\Delta \lambda = 35.0$ nm,
\item[(f)]$\lambda_0 = 657.1$ nm and $\Delta \lambda = 8.6$ nm,
\end{itemize}
\item A tip-tilt mirror and a Boston Micromachines deformable mirror of 32x32 actuators (only 27x27 used)
\item A coronagraph including:
\item[(a)] An FQPM, optimized for the wavelength $\lambda_{opt} = 637$ nm
\item[(b)] A Lyot stop with a diameter of 8 mm for an entrance pupil of 8.1 mm
\item[(c)] In this same plane a reference pupil of variable size (from 0.3 mm to 2 mm). A set of motors allows to choose between Lyot pupil or the reference pupil only. 
\item An sCMOS camera with a readout noise of 1.3 electrons and a full well capacity of 56,000 electrons
\item A set of neutral density filters used to record unsaturated non-coronagraphic images to measure contrast level.
\end{enumerate}
The bench is located in a thermostatically controlled room and is hooded to limit air turbulence between optics. 

\section{Experimental results} 
\label{sec:results}

In this section, we present the experimental results obtain on the THD optical bench in monochromatic (Section~\ref{sec:monochromatic_res}) and polychromatic light (Section~\ref{sec:polychromatic_res}). To measure the achieved contrast inside the DH, we normalize by the maximum of the point spread functions (PSF) obtained without coronagraph. Then we plot the radial profiles of the azimuthal standard deviation (in RMS) of the intensities in the focal plane as a function of the distance to the star (measure in $\lambda_0/D$). We exclude from this measurement the zones located less than 2 $\lambda_0/D$ away from the FQPM transitions. Indeed, the light coming from a planet located on a transition would suffer a strong deformation and a weak transmission.

\subsection{Monochromatic light} 
\label{sec:monochromatic_res}

\begin{figure}
 \begin{center}
 \begin{tabular}{cc}
 \includegraphics[height=5cm]{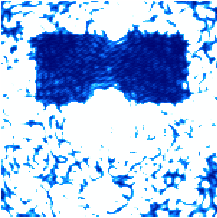}
 \includegraphics[height=8cm]{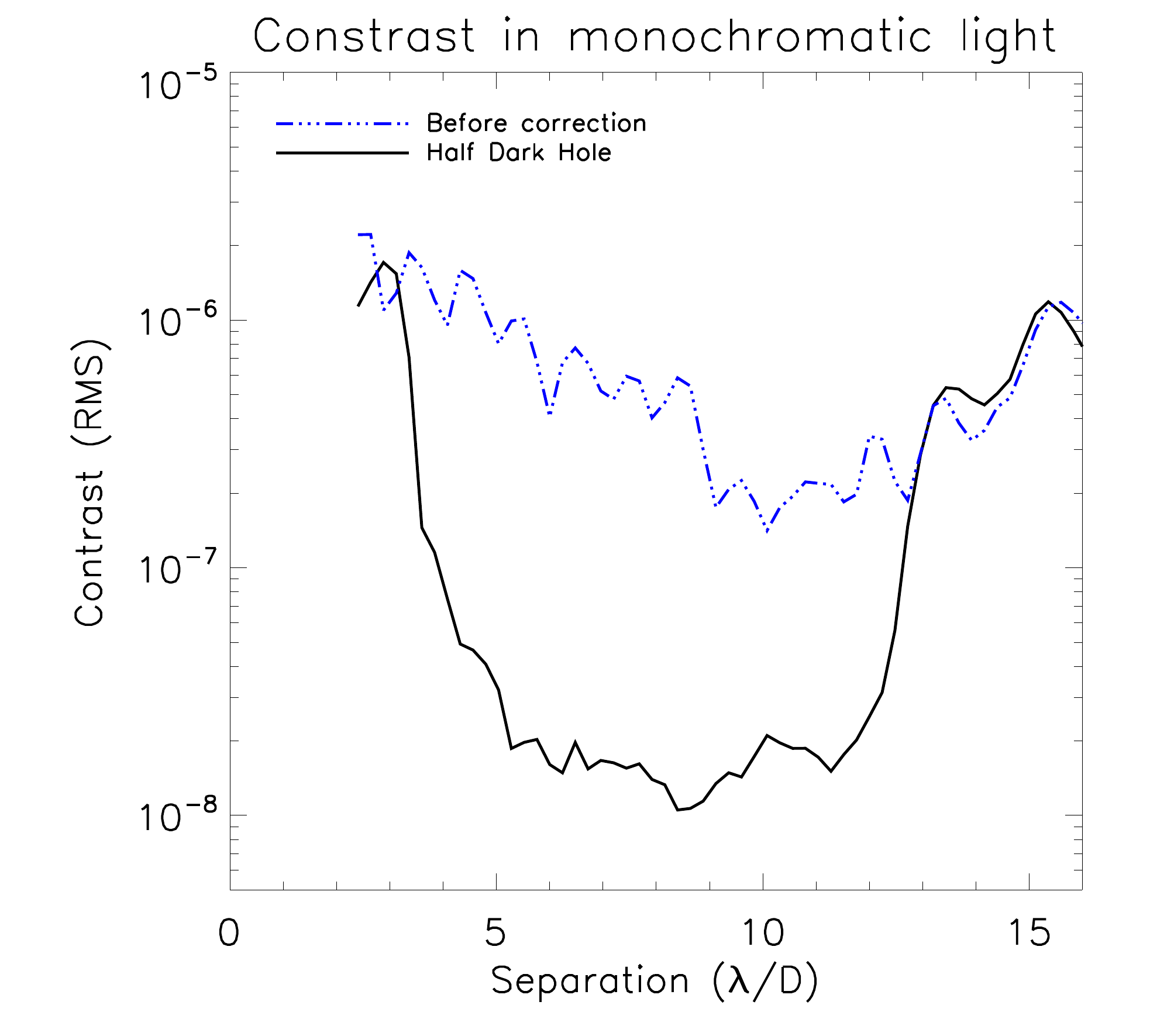}
 \end{tabular}
 \end{center}
 \caption[res_mono] 
{ \label{fig:res_mono} 
On the left, a DH obtained for monochromatic light at 637 nm. Dark zones correspond to high contrasts. On the right, the radial profiles of the azimuthal standard deviation (in RMS) of the intensities in the focal plane normalized by the maximum of the PSF without coronagraph. The dashed-blue line is the result before the correction by the DM and the black-solid line is a typical result obtained after a few iterations.}
\end{figure}

We present on a Figure~\ref{fig:res_mono} a DH obtain with a monochromatic light at the optimum wavelength of our FQPM coronagraph (637 nm). Because, we want to correct phase and amplitude defects with only one DM, we limit the correction zone to an half-DH.
On the left, the radial profiles of the azimuthal standard deviation (in RMS) of the intensities in the focal plane normalized by the maximum of the PSF without coronagraph. The dashed-blue line is the result before the correction by the DM and the black-solid line is a typical result obtain after a few iterations. We obtain contrast level of $1.10^{-7}$ between 4 and 13 $\lambda/D$ and even $2.10^{-8}$ between 5 and 12 $\lambda/D$. We are currently limited by the high amplitude defects introduced by our DM. For more details about limitations in monochromatic light, see Galicher et al. (2013)\cite{galicher13_spie}.

\subsection{Polychromatic light} 
\label{sec:polychromatic_res}

\begin{figure}
 \begin{center}
 \begin{tabular}{cc}
 \includegraphics[height=7.8cm]{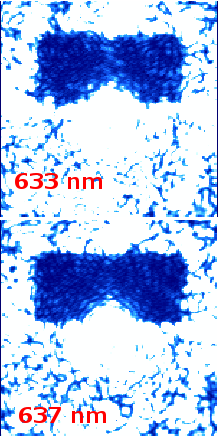}
 \includegraphics[height=7.8cm]{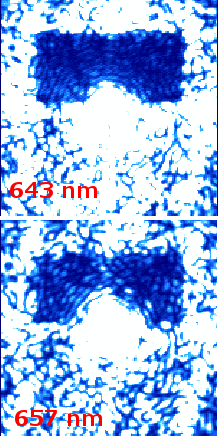}
 \includegraphics[height=7.8cm]{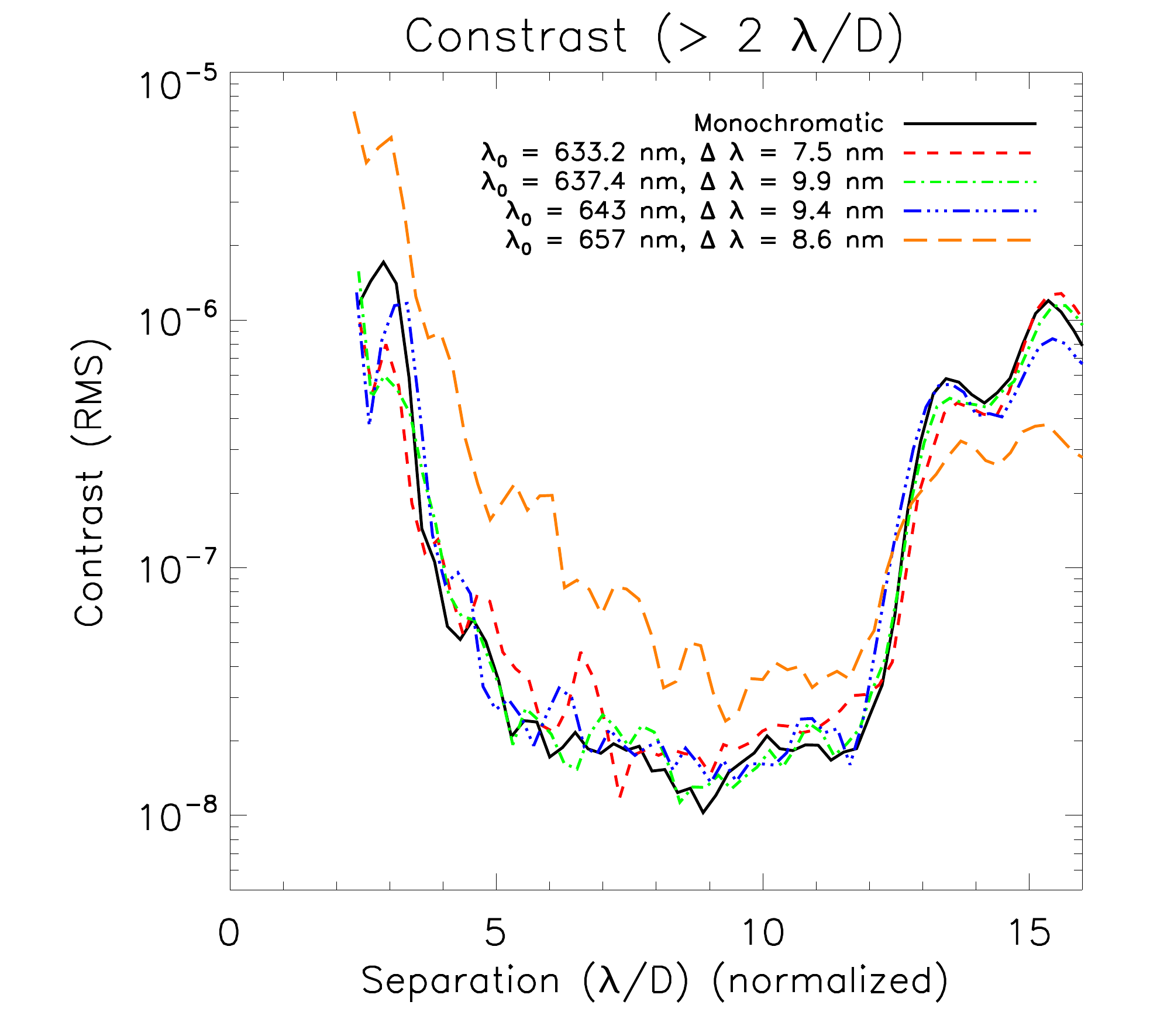}
 \end{tabular}
 \end{center}
 \caption[res_poly_narr] 
{ \label{fig:res_poly_narr} 
On the left, DHs obtained for polychromatic light at $\lambda_0 = 633.2$ nm ($\Delta \lambda = 7.5$ nm), $\lambda_0 = 637.4$ nm ($\Delta \lambda = 9.9$ nm), $\lambda_0 = 643.7$ nm ($\Delta \lambda = 9.4$ nm), $\lambda_0 = 657.1$ nm ($\Delta \lambda = 8.6$ nm). Dark zones correspond to high contrasts. On the right, radial profiles of the azimuthal standard deviation (in RMS) of the intensities in the focal plane normalized by the maximum of the PSF without coronagraph for all these bandwidths and for monochromatic light at 637 nm (black-solid). The abscissa is a re-scaled $\lambda/D$ to obtain same size DHs.}
\end{figure}

\begin{figure}[b]
 \begin{center}
 \includegraphics[height=6cm]{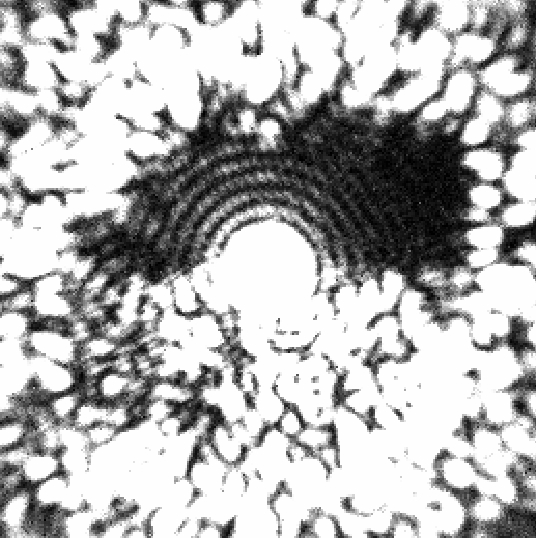}
 \end{center}
 \caption[res_poly_broad] 
{ \label{fig:res_poly_broad}  
DHs obtained in broad band (35 nm around 652 nm)}
\end{figure}

In this section we present the results in polychromatic light. We differentiate between narrow band (Figure~\ref{fig:res_poly_narr}) and broad band (Figure~\ref{fig:res_poly_broad}) results. In Figure~\ref{fig:res_poly_narr}, the DH for all the narrow filters are presented on the left. Dark zones correspond to high contrasts. To obtain these corrections, we created an interaction matrix using the method described in Section~\ref{sec:matrix}. On the right, we represent radial profiles of azimuthal standard deviation (in RMS) of the intensities in the focal plane normalized by the maximum of the PSF without coronagraph for all these filters and for monochromatic light at 637 nm (black-solid). The abscissa is a re-scaled $\lambda/D$ for each wavelength to obtain same size DHs. The results in narrow band do not differ strongly from the monochromatic case, except for the filter with $\lambda_{0} = 657$ nm. The optimal wavelength of our FQPM coronagraph is $\lambda_{opt} = 637$ nm. As described in Galicher et al. 2011\cite{Galicher11}, for $\lambda \neq \lambda_{opt}$, the on-axis light is not diffracted outside the geometrical pupil by the FQPM and the stellar extinction is not perfect. Therefore, for wavelengths distant from $\lambda_{opt} = 637$ a PSF appears into the focal plane. This PSF is not created by wavefront errors and therefore is beyond correction by the DM. Our results in narrow band are therefore currently limited by the chromaticity of the FQPM that we are using. 

For larger broad band, as described in Section~\ref{sec:polychromatic_light_theo}, the fringes are blurred and do not cover the entire DH. We used a 35 nm broad filter, centered around 652 nm. Figure~\ref{fig:res_poly_broad}  shows an half DH. Only the fringed speckles are estimate and corrected and the DH is shrunk. Even more than in narrow band, the results in contrast are highly limited by the monochromatic FQPM. We can distinctly see the PSF inside the corrected DH due to the chromaticity of the coronagraphic mask we are using. 

\section{Conclusion} 
\label{sec:concl}

We demonstrated that the SCC associated with a FQPM produces contrast level of $2.10^{-8}$ between 5 and 12 $\lambda/D$ on an experimental bench. We believe that we are currently limited by high amplitude defects produce by the surface of our DM. In narrow band polychromatic band around the optimum wavelength of our chromatic coronagraph, we prove that we were able to close the loop and perform a correction with the same performance than in monochromatic light. We also analyzed the limitation of the SCC in broad band polychromatic light (blurring of the fringes far away from the zeroth order) but we might be able to overcome this limitation by the introduction of a second reference pupil. The results in contrast are limited by the chromaticity of our FQPM. To enhanced these results in polychromatic light, new coronagraphs are currently under integration on the bench, such as the multi four quadrant phase mask coronagraph\cite{Galicher11} or the the dual zone phase mask coronagraph\cite{Ndiaye12}.


\bibliography{bibliSPIE}   

\begin{thebibliography}{10}

\bibitem{Marois06}
C.~{Marois}, D.~{Lafreni{\`e}re}, B.~{Macintosh}, and R.~{Doyon}, ``{Accurate
  Astrometry and Photometry of Saturated and Coronagraphic Point Spread
  Functions},'' {\em ApJ}~{\bf 647}, pp.~612--619, Aug. 2006.

\bibitem{Lagrange_Bpic09}
A.~{Lagrange}, M.~{Kasper}, A.~{Boccaletti}, G.~{Chauvin}, D.~{Gratadour},
  T.~{Fusco}, D.~{Ehrenreich}, D.~{Apai}, D.~{Mouillet}, and D.~{Rouan},
  ``{Constraining the orbit of the possible companion to {$\beta$} Pictoris.
  New deep imaging observations},'' {\em Astronomy and Astrophysics}~{\bf 506},
  pp.~927--934, Nov. 2009.

\bibitem{Beuzit08}
J.-L. {Beuzit}, M.~{Feldt}, K.~{Dohlen}, D.~{Mouillet}, P.~{Puget}, F.~{Wildi},
  L.~{Abe}, J.~{Antichi}, A.~{Baruffolo}, P.~{Baudoz}, A.~{Boccaletti},
  M.~{Carbillet}, J.~{Charton}, R.~{Claudi}, M.~{Downing}, C.~{Fabron},
  P.~{Feautrier}, E.~{Fedrigo}, T.~{Fusco}, J.-L. {Gach}, R.~{Gratton},
  T.~{Henning}, N.~{Hubin}, F.~{Joos}, M.~{Kasper}, M.~{Langlois}, R.~{Lenzen},
  C.~{Moutou}, A.~{Pavlov}, C.~{Petit}, J.~{Pragt}, P.~{Rabou}, F.~{Rigal},
  R.~{Roelfsema}, G.~{Rousset}, M.~{Saisse}, H.-M. {Schmid}, E.~{Stadler},
  C.~{Thalmann}, M.~{Turatto}, S.~{Udry}, F.~{Vakili}, and R.~{Waters},
  ``{SPHERE: a planet finder instrument for the VLT},'' in {\em Society of
  Photo-Optical Instrumentation Engineers (SPIE) Conference Series},  {\em
  Society of Photo-Optical Instrumentation Engineers (SPIE) Conference Series}
  {\bf 7014}, Aug. 2008.

\bibitem{Macintosh08}
B.~A. {Macintosh}, J.~R. {Graham}, D.~W. {Palmer}, R.~{Doyon}, J.~{Dunn}, D.~T.
  {Gavel}, J.~{Larkin}, B.~{Oppenheimer}, L.~{Saddlemyer},
  A.~{Sivaramakrishnan}, J.~K. {Wallace}, B.~{Bauman}, D.~A. {Erickson},
  C.~{Marois}, L.~A. {Poyneer}, and R.~{Soummer}, ``{The Gemini Planet Imager:
  from science to design to construction},'' in {\em Society of Photo-Optical
  Instrumentation Engineers (SPIE) Conference Series},  {\em Society of
  Photo-Optical Instrumentation Engineers (SPIE) Conference Series} {\bf 7015},
  July 2008.

\bibitem{Boccaletti05_MIRI}
A.~{Boccaletti}, P.~{Baudoz}, J.~{Baudrand}, J.~M. {Reess}, and D.~{Rouan},
  ``{Imaging exoplanets with the coronagraph of JWST/MIRI},'' {\em Advances in
  Space Research}~{\bf 36}, pp.~1099--1106, 2005.

\bibitem{Baudoz06}
P.~{Baudoz}, A.~{Boccaletti}, J.~{Baudrand}, and D.~{Rouan}, ``{The
  Self-Coherent Camera: a new tool for planet detection},'' in {\em IAU Colloq.
  200: Direct Imaging of Exoplanets: Science and Techniques},  {C.~Aime and
  F.~Vakili}, ed., pp.~553--558, 2006.

\bibitem{Galicher10}
R.~{Galicher}, P.~{Baudoz}, G.~{Rousset}, J.~{Totems}, and M.~{Mas},
  ``{Self-coherent camera as a focal plane wavefront sensor: simulations},''
  {\em Astronomy and Astrophysics}~{\bf 509}, pp.~A31+, Jan. 2010.

\bibitem{mazoyer13}
J.~{Mazoyer}, P.~{Baudoz}, R.~{Galicher}, M.~{Mas}, and G.~{Rousset},
  ``{Estimation and correction of wavefront aberrations using the self-coherent
  camera: laboratory results},'' {\em Astronomy and Astrophysics}~{\bf 557},
  p.~A9, Sept. 2013.

\bibitem{galicher13_spie}
R.~{Galicher}, G.~{Rousset}, P.~{Baudoz}, and J.~{Mazoyer}, ``{High-contrast
  imaging with a self-coherent camera},'' in {\em Society of Photo-Optical
  Instrumentation Engineers (SPIE) Conference Series},  {\em Society of
  Photo-Optical Instrumentation Engineers (SPIE) Conference Series}, 2013.

\bibitem{Galicher11}
R.~{Galicher}, P.~{Baudoz}, and J.~{Baudrand}, ``{Multi-stage four-quadrant
  phase mask: achromatic coronagraph for space-based and ground-based
  telescopes},'' {\em Astronomy and Astrophysics}~{\bf 530}, p.~A43, June 2011.

\bibitem{Ndiaye12}
M.~{N'diaye}, K.~{Dohlen}, S.~{Cuevas}, R.~{Soummer},
  C.~{S{\'a}nchez-P{\'e}rez}, and F.~{Zamkotsian}, ``{Improved achromatization
  of phase mask coronagraphs using colored apodization},'' {\em Astronomy and
  Astrophysics}~{\bf 538}, p.~A55, Feb. 2012.

\end{thebibliography}
\bibliographystyle{spiebib}   
\end{document}